\documentclass[conference,10pt]{IEEEtran}
\usepackage{graphicx} %
\usepackage{tikz}
\usepackage{stfloats}
\usepackage[table]{xcolor}
\usepackage{tabularray}
\UseTblrLibrary{booktabs}
\UseTblrLibrary{varwidth} 
\usetikzlibrary{shapes.geometric, shadows.blur, arrows.meta, calc}
\usepackage[colorlinks=true,urlcolor=teal,
            linkcolor=teal,citecolor=teal]{hyperref}
\usepackage{layouts}
\usepackage[protrusion,expansion]{microtype}
\usepackage[backend=biber,style=ieee,maxnames=3,citetracker=true,citestyle=ieee-comp,defernumbers=true]{biblatex}
\usepackage{orcidlink}

\DeclareBibliographyCategory{showinbib}

\newbibmacro*{title+link}[1]{%
  \iffieldundef{doi}
    {%
      \iffieldundef{url}
        {#1}%
        {\href{\thefield{url}}{\textcolor{teal}{#1}}}%
    }
    {\href{https://doi.org/\thefield{doi}}%
       {\textcolor{teal}{#1}}}%
}

\DeclareFieldFormat
  [online,article,inbook,incollection,inproceedings,patent,thesis,unpublished,misc]
  {title}{%
    \usebibmacro{title+link}{#1}%
  }

\renewbibmacro*{doi+eprint+url}{}
\renewbibmacro*{url+urldate}{}

\AtEveryBibitem{
	\clearfield{pages}
	\clearfield{isbn}
	\clearfield{eprint}
	\clearfield{issn}
	\clearfield{abstract}
	\clearfield{volume}
	\clearfield{number}
	\clearfield{month}
}

\usepackage{makecell}
\usepackage{amsmath}
\usepackage[T1]{fontenc}
\usepackage{newtxtext}
\usepackage{newtxmath}
\usepackage[capitalise]{cleveref}

\usepackage{fontawesome5}
\usepackage{pdfrender}

\usetikzlibrary{positioning, calc, backgrounds, arrows.meta, shapes.geometric}

\newcommand{\oicon}[1]{\textpdfrender{TextRenderingMode=1, LineWidth=.5pt}{#1}}
\newcommand{\oiconmask}[1]{\textpdfrender{TextRenderingMode=0, FillColor=panelgray}{#1}}

\definecolor{mypurple}{RGB}{152,128,191}
\definecolor{myorange}{RGB}{240,150,90}
\definecolor{panelgray}{RGB}{237,237,237}
\definecolor{pillgray}{RGB}{142,142,142}

\newcommand{\colfootnote}[1]{\unskip\kern-.1ex\footnote{#1}}

\newcommand{\bookstack}[1]{%
	\begin{scope}[shift={#1}]
		\foreach \dx/\dy/\rot in {-0.17/-0.11/-8, 0/0/0, 0.17/0.11/8}{
				\node[font=\large, text=white, rotate=\rot] at (\dx,\dy) {\oiconmask{\faBook}};
				\fill[panelgray, rotate around={\rot:(\dx,\dy)}]
				(\dx,\dy) circle (0.45em);
				\node[font=\large, text=black!75, rotate=\rot] (book) at (\dx,\dy) {\oicon{\faBook}};
			}
	\end{scope}
}

\newcommand{\personcoding}[2]{%
	\begin{scope}[shift={#1}, xscale=#2]
		\node[font=\Large, text=black!75] (head) at (0,0.30) {\oicon{\faUser}};
		\node[font=\large, text=black!75] (laptop) at (0,-0.16) {\oicon{\faLaptopCode}};
	\end{scope}
}

\newcommand{\folder}[1]{%
	\node[font=\large, text=black!75] at #1 {\oicon{\faFolderOpen}};
}

\newcommand{\pulsesearch}[1]{%
	\begin{scope}[shift={#1}, line width=0.85pt, black]
		\draw (0,0) circle (0.34);
		\draw (-0.24,-0.24) -- (-0.40,-0.40);
		\draw[line width=2.6pt, line cap=round] (-0.40,-0.40) -- (-0.52,-0.52);
		\draw[line cap=round, line join=round]
		(-0.20,0.02) -- (-0.09,0.02) -- (-0.02,0.20) -- (0.07,-0.16) -- (0.13,0.02) -- (0.24,0.02);
	\end{scope}
}

\newcommand{\lengthOf}[1]{
\the#1\\\printinunitsof{cm}\prntlen{#1}\\\printinunitsof{in}\prntlen{#1}\\
}

\newcommand{\generated}[1]{%
	\begingroup
	\microtypesetup{activate=false}
	\IfFileExists{../build/tex/#1.tex}%
		{\input{../build/tex/#1.tex}}%
		{\input{generated/#1.tex}}
	\endgroup
}

\AtEveryCitekey{%
  \ifbool{citetracker}
    {\addtocategory{showinbib}{\thefield{entrykey}}}
    {}%
}

\DeclareCiteCommand{\arxivref}
  {\citetrackerfalse} %
  {%
    \iffieldundef{eprint}
      {\doiid{\thefield{doi}}}
      {\href{https://arxiv.org/abs/\thefield{eprint}}%
             {arXiv:\thefield{eprint}}}%
  }
  {\multicitedelim}
  {}

\newcommand{\doiid}[1]{%
 \href{https://doi.org/#1}{\nolinkurl{#1}}}

\newcommand{\othr}{\textit{Technical University of}\\\textit{Applied Sciences Regensburg}} 
\newcommand{\genemail}[2]{\href{#1}{#2}}

\definecolor{lfdochre}{HTML}{E69F00}
\definecolor{lfdblue}{HTML}{1F78B4}

\title{Repositories, Contributors, and Continuity: An Empirical Study of Foundational Quantum Software}

\author{
\IEEEauthorblockN{Vincent Gierisch\(^{*}\)\orcidlink{0009-0004-0638-6757}}
\IEEEauthorblockA{
    \othr\\
    Regensburg, Germany}
    \genemail{mailto:vincent.gierisch@othr.de}{vincent.gierisch@othr.de}\\
    {\small \(^{*}\)Authors contributed equally}
    \and
\IEEEauthorblockN{Nicole Hoess\(^{*}\)\orcidlink{0009-0002-6194-5887}}
\IEEEauthorblockA{\othr\\
Regensburg, Germany}
\genemail{mailto:nicole.hoess@othr.de}{nicole.hoess@othr.de}
\and
\IEEEauthorblockN{Ralf Ramsauer\(^{*}\)\orcidlink{0009-0007-4033-8515
}}
\IEEEauthorblockA{\othr\\
Regensburg, Germany}
\genemail{mailto:ralf.ramsauer@othr.de}{ralf.ramsauer@othr.de}
\and
    \IEEEauthorblockN{Wolfgang Mauerer\orcidlink{0000-0002-9765-8313}}
    \IEEEauthorblockA{\othr\\
        \textit{Siemens AG, Technology} \\
        Regensburg/Munich, Germany \\
\genemail{mailto:wolfgang.mauerer@othr.de}{wolfgang.mauerer@othr.de}
    }
}

\begin{document}
\maketitle

\begin{abstract}
Driven by contributions from academia, industry, and open-source communities, the quantum software ecosystem is rapidly growing.
Across this ecosystem, new concepts often emerge through software artefacts accompanying scientific publications as well as through sustained development in larger communities.
However, many repositories receive development efforts only over a limited period of time, raising the question whether their concepts persist beyond individual repositories.

In this paper, we apply established empirical software engineering techniques to analyse a set of foundational quantum software repositories.
We combine cross-repository activity with contributor relationships to study the evolution of communities. %
Our analysis provides empirical evidence of contributor migration patterns and indications of cross-project knowledge transfer.
We observe multiple development paths: projects may evolve into sustained communities, contributors may
integrate concepts 
into established ecosystems, or activity may continue through new and follow-up software artefacts.
Our observations provide an initial empirical perspective on how concepts and influence persist across repository boundaries in quantum software ecosystems.
\end{abstract}


\section{Introduction}
Quantum computing has led to the emergence of a rapidly growing software ecosystem that comprises applications, high-level programming abstractions, compiler stacks, intermediate representations, as well as software for pulse execution and control.
Research-driven developments play a central role in this ecosystem.
Frequently, scientific 
publications introduce new concepts through accompanying software 
artefacts that serve as standalone proof of concept implementations.
At the same time, larger community-driven efforts contribute to sustained \emph{long-term maintained} software.
However, many repositories receive development effort only over a limited period of time.
This raises the question if concepts of \emph{short-lived} projects are transferred into long-term maintained development efforts.

In the field of software engineering (SE), empirical analysis of software ecosystems has long been established as a means to study repository evolution~\cite{joblin_evolutionary_2017}, project activity~\cite{ait_empirical_2022}, community dynamics~\cite{joblin_developer_2015}, and long-term sustainability~\cite{ramsauer2019list}.
Within SE, Mining Software Repositories (MSR)~\cite{hemmati2013msr} techniques provide perspectives that extend beyond functional or performance-oriented evaluation.
Such perspectives are also beginning to gain traction in quantum computing~\cite{murillo_quantum_2025}, with recent work~\cite{upadhyay2025analyzing,ramalho2026mining,yousuf2025characterizing,destefano_qse_2024,khan_mining_2025} studying repository evolution and maintenance, software patterns and quality characteristics in quantum projects.

While these developments establish MSR as an emerging perspective for quantum software, existing analyses primarily characterise isolated repositories and their observable development artefacts.
Less attention has been paid to relationships \emph{across} repositories and to whether the influence of projects on the ecosystem persists beyond periods of active development.

In this work, we combine repository activity with contributor relationships across repositories to analyse whether and how development activity and influence extend beyond individual projects.
Therefore, we study a subset of foundational repositories\footnote{Related publications:
\arxivref{mccaskey_qcor_2021},
\arxivref{mccaskey_mlir_2021}, %
\arxivref{kim_cudaq_2023}, %
\arxivref{burgholzer_mqtcore_2025}, %
\arxivref{burgholzer_26_mqtcc}, %
\arxivref{ittah_catalyst_2024}, %
\arxivref{bergholm_pennylane_2022},
\arxivref{ittah_qiro_2021},
\arxivref{svore_qdk_2018}, %
\arxivref{javadi_qiskit_2024},
\arxivref{healy_qecompiler_pre_2024},
\arxivref{koch_hugr_2025},
\arxivref{peduri_qssa_2022},
\arxivref{khammassi_openql_2020},
\arxivref{wong_qiree_2025},
\arxivref{adams_qwerty_2024}.
}
from quantum computing compiler projects 
spanning research-driven and community-driven development.

Our analysis provides empirical evidence of contributor migration patterns and indications of cross-project transfer and continuation of concepts.
We observe multiple development paths:
  repositories may evolve into sustained communities,
  contributors may carry development efforts into established ecosystems,
  or development may continue through new and follow-up software artefacts.
This represents an encouraging signal for quantum software researchers: although cross-project knowledge transfer can still be improved, the first pathways for research to make an impact already exist.
Overall, we make the following contributions:
\begin{itemize}
\item We construct a dataset of foundational quantum software repositories and analyse their development dynamics using established empirical software engineering techniques.
\item We provide a network representation of the quantum ecosystem to quantify contributor relationships across repositories and identify migration patterns between research-driven and sustained community ecosystems.
\item We show indications that repository inactivity does not necessarily imply disappearing influence and discuss cross-project continuation beyond the original repository.
\item We provide a self-contained reproduction package~\cite{mauerer_beyond_2022, mauerer_1-2-3_2022} with all data, scripts and the execution environment.\colfootnote{\url{https://github.com/lfd/qset26-reproduction}, \href{https://doi.org/10.5281/zenodo.21628842}{10.5281/zenodo.21628842}}
\end{itemize}

\section{Methodology}
\begin{figure*}[t!]
    \centering
    \resizebox{\textwidth}{!}{%
        \begin{tikzpicture}[
	>={Stealth[length=2.2mm]},
	font=\sffamily,
	pill/.style       = {rectangle, rounded corners=2pt, fill=pillgray,
			text=white, font=\bfseries, align=center,
			inner sep=4pt, minimum height=0.55cm, anchor=north west},
	pillcolor/.style  = {rectangle, rounded corners=2pt, text=white,
			font=\bfseries, align=center, inner sep=4pt,
			minimum height=0.55cm, anchor=north west},
	stepcircle/.style = {circle, draw=black, inner sep=0pt,
			minimum size=0.7cm, font=\large},
	steptitle/.style  = {font=\large, anchor=west, yshift=-1pt},
	panel/.style      = {rounded corners=4pt, fill=panelgray, draw=none},
	bigicon/.style     = {font=\Large, text=black!75},
	]

	\node[stepcircle] (s1) at (0,4.2) {1};
	\node[steptitle, right=6pt of s1] (title1) {Select quantum subject projects};

	\node[panel, minimum width=4cm, minimum height=2.5cm, anchor=north west] (panel-commits) at (0,3.6) {};
	\node[pill, minimum width=4cm] (lbl-commits) at (panel-commits.north west) {Commits};

	\begin{scope}[shift={($(panel-commits.center)+(-2cm,-0.75cm)$)},
		roundnode/.style={circle, draw=black, fill=myorange!70, inner sep=2.2pt},
		arr/.style={-{Stealth[length=1.6mm]}, thick}]
		\foreach \i [count=\xi] in {1, ..., 7}
		\node[roundnode] (round\i) at (\xi*0.5cm,0.5cm) {};

		\foreach \i [count=\xi] in {1,2}
		\node[roundnode] (roundsecond\i) at (0.5cm+\xi*0.5cm,0cm) {};

		\foreach \i [count=\xi] in {1,2}
		\node[roundnode] (roundthird\i) at (1.5cm+\xi*0.5cm,1cm) {};

		\foreach \i [evaluate=\i as \nexti using int(\i+1)] in {1,2,3,4,5,6}
		\draw [->] (round\i) -- (round\nexti);

		\foreach \i [evaluate=\i as \nexti using int(\i+1)] in {1}
		\draw [->] (roundsecond\i) -- (roundsecond\nexti);

		\foreach \i [evaluate=\i as \nexti using int(\i+1)] in {1}
		\draw [->] (roundthird\i) -- (roundthird\nexti);

		\draw [->] (round1) -- (roundsecond1);
		\draw [->] (roundsecond2) -- (round4);
		\draw [->] (round3) -- (roundthird1);
		\draw [->] (roundthird2) -- (round6);
	\end{scope}

	\node[panel, minimum width=3.5cm, minimum height=2.5cm, anchor=north west]
	(panel-pubs) at ($(panel-commits.north east)+(0.35,0)$) {};
	\node[pill, minimum width=3.5cm] (lbl-pubs) at (panel-pubs.north west) {Publications};

	\bookstack{($(panel-pubs.center)+(-1.0,-0.40)$)}
	\bookstack{($(panel-pubs.center)+( 0.05,0.05)$)}
	\bookstack{($(panel-pubs.center)+( 1.0,-0.40)$)}

	\coordinate (step2-x0) at ($(panel-pubs.north east)+(1.25,0)$);
	\node[stepcircle] (s2) at (step2-x0 |- s1) {2};
	\node[steptitle, right=6pt of s2] (title2) {Aggregate and analyse repository data};

	\node[panel, minimum width=4.2cm, minimum height=2.5cm, anchor=north west]
	(panel-cross) at (step2-x0 |- panel-commits.north) {};
	\node[pill, minimum width=4.2cm] (lbl-cross) at (panel-cross.north west) {Cross-project mapping};

	\folder{($(panel-cross.center)+(-1.60,-0.20)$)}
	\personcoding{($(panel-cross.center)+(-0.85,-0.40)$)}{1}
	\node[font=\large, text=black!75] at ($(panel-cross.center)+(0,0.1)$) {\oicon{\faClock}};
	\personcoding{($(panel-cross.center)+(0.85,-0.40)$)}{-1}
	\folder{($(panel-cross.center)+(1.60,-0.20)$)}
	\draw[<->, thick, black!70]
	($(panel-cross.center)+(-0.45,-0.45)$) -- ($(panel-cross.center)+(0.45,-0.45)$);

	\node[panel, minimum width=3.4cm, minimum height=2.5cm, anchor=north west]
	(panel-eco) at ($(panel-cross.north east)+(0.3,0)$) {};
	\node[pill, minimum width=3.4cm] (lbl-eco) at (panel-eco.north west) {Ecosystem view};

	\begin{scope}[shift={($(panel-eco.center)+(0.12,-0.2)$)}, scale=0.7,
			netnode/.style={circle, draw=black, fill=white, line width=0.55pt, inner sep=1.3pt},
			netedge/.style={black, line width=0.5pt}]
		\node[netnode, minimum size=0.42cm] (hub1) at (-1.05,-0.15) {};
		\foreach \ang/\rad in {95/0.62, 155/0.55, 205/0.6, 250/0.58, 330/0.55}
		\node[netnode] (l-\ang) at ($(hub1)+(\ang:\rad)$) {};
		\foreach \ang/\rad in {95/0.62, 155/0.55, 205/0.6, 250/0.58, 330/0.55}
		\draw[netedge] (hub1) -- (l-\ang);
		\node[netnode] (c1) at (-0.35,-0.15) {};
		\node[netnode] (c2) at (0.10,-0.20) {};
		\draw[netedge] (hub1) -- (c1) -- (c2);
		\node[netnode, minimum size=0.34cm] (hub2) at (0.75,0.0) {};
		\foreach \ang/\rad in {15/0.55, 75/0.6, 130/0.58, 270/0.55, 330/0.55}
		\node[netnode] (r-\ang) at ($(hub2)+(\ang:\rad)$) {};
		\draw[netedge] (c2) -- (hub2);
		\foreach \ang/\rad in {15/0.55, 75/0.6, 130/0.58, 270/0.55, 330/0.55}
		\draw[netedge] (hub2) -- (r-\ang);
		\node[netnode] (mol1) at (-0.30,-0.95) {};
		\node[netnode] (mol2) at (0.05,-0.95) {};
		\draw[netedge] (mol1) -- (mol2);
	\end{scope}

	\node[panel, minimum width=3.0cm, minimum height=2.5cm, anchor=north west]
	(panel-temp) at ($(panel-eco.north east)+(0.3,0)$) {};
	\node[pill, minimum width=3.0cm] (lbl-temp) at (panel-temp.north west) {Temporal view};

	\begin{scope}[shift={($(panel-temp.center)+(0,-0.265)$)}, scale=0.7]
		\fill[white] (-0.88,-1.085) rectangle (0.88,1.085);
		\draw[black!70, thick] (-0.88,-1.085) rectangle (0.88,1.085);
		\draw[myorange, thick]
		(-0.88,-0.174) -- (-0.563,-0.347) -- (-0.176, 0.174) --
		( 0.00, 0.00) -- ( 0.335, 0.00) -- ( 0.528,-0.152) -- ( 0.88, 0.347);
		\draw[black!45, thick]
		(-0.88,-0.360) -- (-0.563,-0.533) -- (-0.176,-0.012) --
		( 0.00,-0.186) -- ( 0.335,-0.186) -- ( 0.528,-0.338) -- ( 0.88, 0.161);
	\end{scope}

	\begin{scope}[scale=0.8]
      \pulsesearch{($(panel-eco.south east)+(0.2,1.5)$)}
    \end{scope}

	\coordinate (step3-x0) at ($(panel-temp.north east)+(1.25,0)$);
	\node[stepcircle] (s3) at (step3-x0 |- s1) {3};
	\node[steptitle, right=6pt of s3] (title3) {Interpret results};

	\node[panel, minimum width=3.6cm, minimum height=2.5cm, anchor=north west]
	(panel-insights) at (step3-x0 |- panel-commits.north) {};
	\node[pill, minimum width=3.6cm] (lbl-insights) at (panel-insights.north west) {Insights};

	\begin{scope}[shift={($(panel-insights.center)+(0, -0.25)$)}]

		\bookstack{(-1,0.25)}

		\draw[<->, thick, black!70]
		(-0.4,0.25) -- (0.45,0.25);

		\node[bigicon, font=\huge] at (1,0.25) {\oicon{\faCubes}};

		\node[bigicon] at (0.05,-0.25) {\oicon{\faLightbulb}};

	\end{scope}

	\coordinate (sep1-x) at ($(panel-pubs.east)!.5!(panel-cross.west)$);
	\coordinate (sep2-x) at ($(panel-temp.east)!.5!(panel-insights.west)$);
	\coordinate (seps-top) at (0,3.6);
	\coordinate (seps-bot) at (panel-cross.south);
	\draw[dashed, gray!55, line width=0.9pt] (sep1-x |- seps-top) -- (sep1-x |- seps-bot);
	\draw[dashed, gray!55, line width=0.9pt] (sep2-x |- seps-top) -- (sep2-x |- seps-bot);

    \draw[->, thick, black!70] (title1.east) -- (sep1-x |- title1.east);
    \draw[->, thick, black!70] (title2.east) -- (sep2-x |- title2.east);

\end{tikzpicture}
    }
    \vspace*{-1.2em}
    \caption{Overview of the analysis workflow. We first relate quantum software repositories to associated publications and derive repository-level observations from the development history. We then map contributor identities across repositories and construct complementary ecosystem and temporal views to capture relations and evolution beyond individual projects. This allows for analysing activity, contributor dynamics, and the continuation of concepts in the ecosystem.}
    \label{fig:method}
\end{figure*}

\begin{figure}[b]
    \centering
    \vspace{-1.1em}
    \input{../build/tex/network_plot.tex}
    \vspace{-1.9em}
    \caption{Overview of the quantum software ecosystem with projects (grey nodes) and contributors (blue nodes and edges for developers contributing to multiple projects; lilac nodes and edges for developers contributing to a single project). Developer nodes are scaled by their total commits across the entire ecosystem. Edge widths indicate the share of commits (effort) a developer contributed to the specific project during their active months.\label{fig:network}}
\end{figure}

\begin{table}[t]
  \centering
  \scriptsize
  \caption{Subject project statistics and ecosystem metrics.}
  \label{tab:projects}
    \begin{tblr}{colspec={Xrrrlrrrrrr},
                colsep=2.9pt,rowsep=0.6pt,
                row{odd}={bg=gray!15},
                row{1}={bg=white,fg=black}}
        \toprule
        Project & \rotatebox{90}{\shortstack[l]{Commits}} & \rotatebox{90}{\shortstack[l]{LoC[k]}} & \rotatebox{90}{\shortstack[l]{Team Size}} & \rotatebox{90}{\shortstack[l]{Active\\Development\\Time}} & \rotatebox{90}{\shortstack[l]{Exclusive\\Dev. Ratio}} & \rotatebox{90}{\shortstack[l]{Cross-Project\\Dev. Ratio}} & \rotatebox{90}{\shortstack[l]{Core Devs.}} & \rotatebox{90}{\shortstack[l]{Peripheral\\Devs.}} & \rotatebox{90}{\shortstack[l]{Median\\Projects/Dev.}} & \rotatebox{90}{\shortstack[l]{Linked\\Projects}} \\ \midrule
        \textsc{catalyst} & 2,306 & 156 & 62 & 2023--2026 & 0.21 & 0.79 & 13 & 49 & 2.00 & 5 \\
        \textsc{cuda-q} & 2,777 & 310 & 105 & 2023--2026 & 0.81 & 0.19 & 10 & 95 & 1.00 & 12 \\
        \textsc{hugr} & 1,719 & 108 & 23 & 2023--2026 & 0.22 & 0.78 & 5 & 18 & 2.00 & 5 \\
        \textsc{mqt-core} & 3,476 & 136 & 45 & 2019--2026 & 0.89 & 0.11 & 4 & 41 & 1.00 & 5 \\
        \textsc{pennylane} & 6,674 & 537 & 223 & 2018--2026 & 0.70 & 0.30 & 25 & 198 & 1.00 & 5 \\
        \textsc{pyqir} & 295 & 19 & 22 & 2021--2026 & 0.18 & 0.82 & 3 & 19 & 2.00 & 11 \\
        \textsc{qat} & 194 & 51 & 8 & 2021--2023 & 0.25 & 0.75 & 2 & 6 & 2.50 & 6 \\
        \textsc{qcor} & 1,217 & 90 & 8 & 2018--2022 & 0.12 & 0.88 & 2 & 6 & 2.00 & 6 \\
        \textsc{qdk} & 2,179 & 624 & 64 & 2023--2026 & 0.80 & 0.20 & 10 & 54 & 1.00 & 8 \\
        \textsc{qe-compiler} & 244 & 28 & 22 & 2022--2024 & 0.59 & 0.41 & 8 & 14 & 1.00 & 2 \\
        \textsc{qiro} & 76 & 7 & 1 & 2020 & 0.00 & 1.00 & 1 & 0 & 4.00 & 3 \\
        \textsc{qiskit} & 9,341 & 363 & 661 & 2017--2026 & 0.93 & 0.07 & 33 & 628 & 1.00 & 11 \\
        \textsc{qiskit-qir} & 78 & 3 & 6 & 2022--2025 & 0.50 & 0.50 & 3 & 3 & 1.50 & 4 \\
        \textsc{qwerty} & 398 & 49 & 13 & 2024--2026 & 0.92 & 0.08 & 1 & 12 & 1.00 & 4 \\
        \textsc{tket2} & 911 & 123 & 30 & 2022--2026 & 0.40 & 0.60 & 4 & 26 & 2.00 & 5 \\
        \textsc{xacc} & 2,140 & 1,580 & 24 & 2016--2023 & 0.58 & 0.42 & 2 & 22 & 1.00 & 8 \\
        \bottomrule
  \end{tblr}

  \vspace{-1.3em}
\end{table}

Our investigation of cross-project influence in the quantum software ecosystem builds on the premise of \emph{developer-mediated knowledge transfer}. A developer who gains expertise within one project may later carry that technical know-how into others~\cite{ma_how_2017, behfar_knowledge_2018}, whether by joining an established community, 
or starting a new initiative. 
Therefore, we hypothesise that the same developer will at some point be an active contributor in multiple projects. To find such knowledge flows, we follow the multi-stage MSR pipeline illustrated in~\cref{fig:method}.

\paragraph{Subject project selection} %
We start with a structured project selection based on expert knowledge to identify a subset of %
software projects related to quantum compilation from academic publications and official repositories.
Projects were included if they provide a publicly accessible implementation of at least one quantum-specific compilation function, while considering diverse properties shown in \cref{tab:projects}, columns 2--4.%

\paragraph{Data extraction and preprocessing} To measure \emph{project activity} as the frequency of commits contributed by developers over the project's lifetime, we extract the development history of each project from the version-control system (VCS) \texttt{git} using the MSR tool \texttt{perceval}~\cite{duenas2018perceval}. %
We filter commits on each project's main branch to exclude development efforts that have not (yet) been integrated.
In addition, we filter bots based on manually identified name patterns and exclude merge commits, as both primarily capture automated or integration events rather than direct development activity.

\paragraph{Cross-project mapping} To analyse developer-induced knowledge flows at an ecosystem level, we must ensure that the same developer can be identified in all projects, even when using different e-mail addresses or nicknames. For identity matching, we consolidate all projects' contributors by merging authors and committers that share one of the following: full name \emph{and} e-mail address, e-mail address, GitHub username or exact name. To ensure correctness, we manually sanity-check the matches.

\enlargethispage{\baselineskip}
\paragraph{Ecosystem network construction} Social networks of open-source software (OSS) communities are an established means to analyse collaboration~\cite{joblin_developer_2015, joblin_classifying_2017, joblin_evolutionary_2017} and knowledge transfer: If a developer of one community is connected to developers of another community, he may contribute knowledge which the other community absorbs~\cite{behfar_knowledge_2018}. Following this premise, we construct an ecosystem network by linking the merged developer identities to all projects they contributed to and calculating ecosystem metrics capturing developer roles (exclusive vs. cross-project developers, core developers responsible for at least 80\% of commits vs. peripheral developers) and structural properties based on links.

\paragraph{Temporal analysis} Based on the static network overview, we analyse the dynamics of cross-project contributions in terms of their temporal order and direction. We count commits per developer, project and day and group results on a monthly basis to visualise cross-project contributor activity over time. This allows for identifying turnover events, in which developers end their activity in one project and possibly join another one. To understand the relationship between turnover, knowledge flows and scientific publication activities, we further annotate our data with project-specific preprint and publication dates collected from digital libraries and repositories. 

\begin{figure*}[t!]
    \input{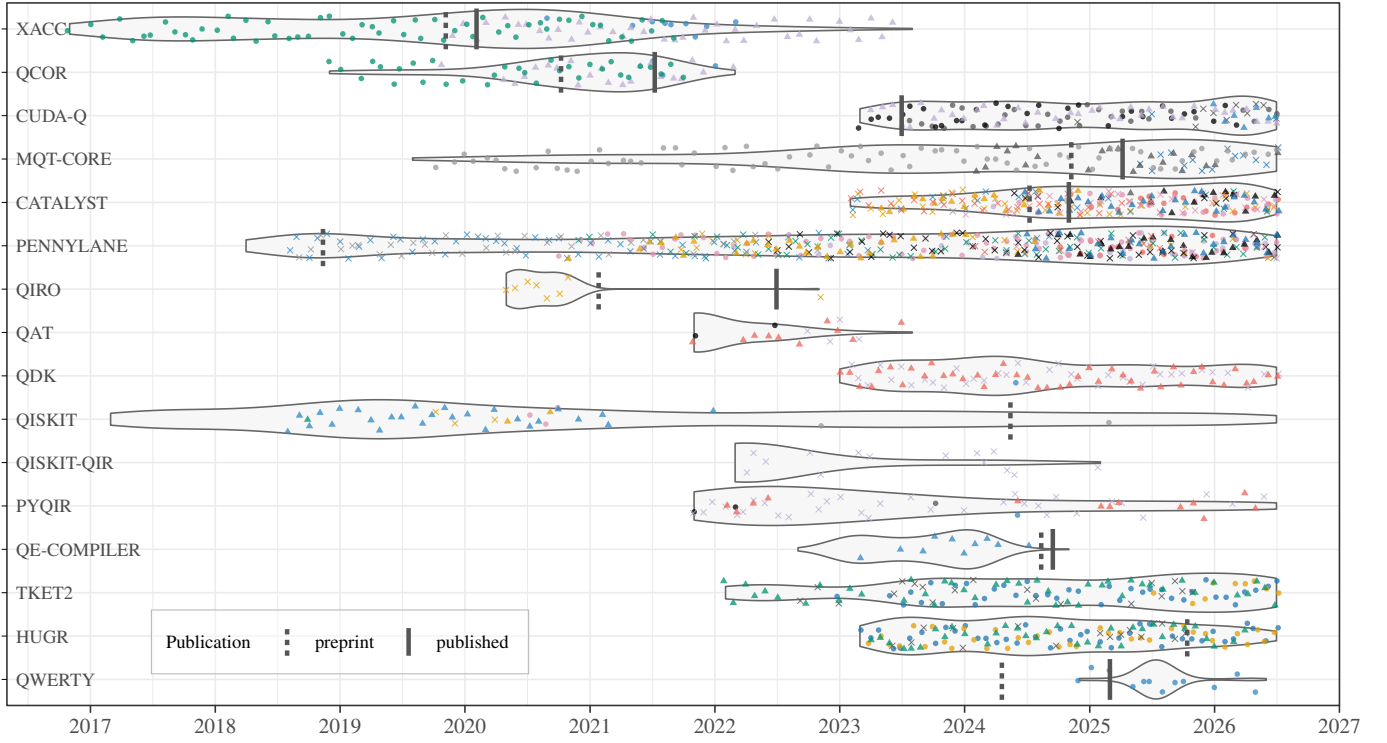}\vspace*{-2em}
    \caption{Contribution activity over time. Violins show the overall commit distribution per project. Coloured points mark the top 20\% (24 of 118 cross-project) contributors by total commits. Each of the colour–shape combinations denotes a single developer, allowing the same contributor to be traced across projects.}
    \label{fig:cross-project-contributions}
\end{figure*}

\section{Results / Evaluation}
\paragraph{Ecosystem structure} \cref{fig:network} shows the ecosystem network with all developers and the projects they contribute to, while \cref{tab:projects} presents complementary quantitative metrics. 
First, in most repositories, the majority of contributors remain active exclusively within their own project.
Despite these largely differing developer bases, no project forms an entirely isolated community but has at least some cross-project contributors, representing candidate connections for knowledge transfer.

Established community-driven projects with an industry backbone, such as \textsc{cuda-q}, \textsc{pennylane}, and \textsc{qiskit}, benefit from their large developer base: more core developers share 80\% of commits, and relative efforts of developers spread more equally. Several connections emerge across projects initiated by different companies, such as Xanadu's \textsc{catalyst}, IBM's \textsc{qiskit} and Microsoft's \textsc{qdk}, possibly indicating joint integration efforts.
Mediated by individual developers, community-driven projects are also connected to academic projects, such as \textsc{qiro}, \textsc{qe-compiler}, \textsc{qcor} and \textsc{xacc}, suggesting that the larger communities support research activities or absorb know-how.

Several short-lived projects, such as \textsc{hugr}, \textsc{qat} and \textsc{qiro}, show activity over only one or two years and were driven by one or a few developers, yet show high ratios of cross-project contributors, suggesting their developers' experience transferred elsewhere.
Connections from these to active projects, as observed for \textsc{mqt-core}, \textsc{pyqir} and \textsc{qwerty}, could indicate parallel developments, collaborations on emerging concepts, follow-up and related research, or integration tasks.
Relatedly, top contributors, by both ecosystem commits and cross-project links, often carry much of the development effort in research projects, while also contributing to industry-initiated ones.

\paragraph{Temporal analysis} The contribution activity in \cref{fig:cross-project-contributions} indicates that repository activity alone does not fully capture the continued influence of research-driven quantum software artefacts.
\cref{fig:cross-project-contributions} provides an indication that know-how in the analysed ecosystem often moves across repository boundaries.
Contributors active in two or more projects connect earlier, smaller repositories with later, larger projects that are often driven by enterprises. 
A comparable overlap is observed between \textsc{xacc} and \textsc{qcor}, which share several contributors. 
This relationship is plausible, as both projects originated from the same research centre \cite{mccaskey_xacc_2019, mccaskey_qcor_2021}. 
Several other prominent contributor overlaps can likewise be explained by shared institutional origins or affiliations. 
This applies to \textsc{catalyst}/\textsc{pennylane}, \textsc{qiskit}/\textsc{qiskit-qir}/\textsc{qe-compiler}, \textsc{hugr}/\textsc{tket2}, and \textsc{qat}/\textsc{pyqir}.
Given the shared institutional origins, knowledge transfer is likely.

Of particular interest is the contributor overlap among \textsc{xacc}, \textsc{qcor}, and \textsc{cuda-q}. The temporal order of contributions suggests that developers previously active in \textsc{xacc} and \textsc{qcor} subsequently contributed to \textsc{cuda-q}. 
This development trajectory indicates that expertise acquired in the earlier projects may have informed subsequent work on \textsc{cuda-q}. A further example of this pattern is the transition from \textsc{qiro} to \textsc{catalyst}, in which activity in the earlier research prototype is followed by contributions to \textsc{catalyst} and \textsc{pennylane} by the same developer.
A manual verification of the commits provides additional support for the interpretation that know-how related to intermediate representations was transferred.
\colfootnote{See \textsc{catalyst} commit \texttt{\href{https://github.com/PennyLaneAI/catalyst/commit/a93629f54f8caa8555a2d05d0c5584d6ba6c96a1}{a93629f}}, \textsc{qcor} commit \texttt{\href{https://github.com/qir-alliance/qcor/commit/28a61e4296ed5dd476463879d938711a83ffbc64}{28a61e4}}, and \textsc{cuda-q} commit \texttt{\href{https://github.com/NVIDIA/cuda-quantum/commit/da88cb5505fde38e7a6465fa7e53b223ca17d060}{da88cb5}}.}

\section{Discussion and Conclusion}
Our study reveals initial patterns of knowledge transfer and sustainability in the quantum ecosystem: evolution into long-term maintenance, novel concepts move into such sustained communities or follow-up activity in new projects. Especially the second pattern is encouraging for researchers. Although the ecosystem is dominated by major vendors, long-term research creates impact by concept integration
into established frameworks, boosted by publishing code and reproduction packages alongside papers. Mauerer et al.~\cite{mauerer_beyond_2022, mauerer_1-2-3_2022} provide actionable guidelines.

Yet many efforts remain within separate communities, which
leaves possibilities to improve knowledge transfer in the quantum ecosystem, as established ecosystems such as Linux demonstrate.
Examining development processes, design abstractions and patterns in depth could identify concepts for cross-project transfer and standardisation, reducing duplicated efforts and accelerating development.

Generalisability of our exploratory study is incomplete: we detect knowledge flows from commit activity without other channels, and rely on MSR techniques with known threats~\cite{hoess_does_2025}.
Additional data and techniques 
would likely reveal further examples rather than undermine the reported results. 

We suggest a follow-up study including developer interviews to recover motivations that technical data cannot fully explain.
We hope that this study provides a starting point for discussions of how knowledge transfer in the quantum software ecosystem currently works, and how it should work in the future.\\

{\footnotesize\noindent\textbf{Acknowledgements}
We thank Lukas Landgraf for his assistance with visualisation.
We acknowledge partial support by the European Regional Development Fund (ERDF) and by the Free State of Bavaria as part of the project AIM-SMEs (Grant No. 2506-014-3.2), co-funded by the European Union; also by the High-Tech Agenda of the Free State of Bavaria, and the German Research Foundation (DFG), grant MA 9739/1-1. 
\par}

\renewbibmacro{in:}{}
\printbibliography[category=showinbib]
\end{document}